# Novel Hierarchical Correlation Functions for Quantitative Representation of Complex Heterogeneous Materials and Microstructural Evolution

Pei-En Chen[1], Wenxiang Xu[2,3], Nikhilesh Chawla[2], Yi Ren[1,*], and Yang Jiao[2,*]

[1] Mechanical Engineering, Arizona State University, Tempe AZ 85287, USA

[2] Materials Science and Engineering, Arizona State University, Tempe AZ 85287, USA

[3] College of Mechanics and Materials, Hohai University, Nanjing 211100, P.R. China

[*] Corresponding author, E-mail: yang.jiao.2@asu.edu (Y. J.) and yiren@asu.edu (Y. R.)

**Abstract**

Effective and accurate characterization and quantification of complex microstructure of a heterogeneous material and its evolution under external stimuli are very challenging, yet crucial to achieving reliable material performance prediction, processing optimization and advanced material design. Here, we address this challenge by developing a set of novel hierarchical statistical microstructural descriptors, which we call the "n-point polytope functions" $P_n$, for quantitative characterization, representation and modeling of complex material microstructure and its evolution. These novel polytope functions successively include higher-order n-point statistics of the features of interest in the microstructure in a concise, expressive, explainable, and universal manner; and can be directly computed from multi-modal imaging data. We develop highly efficient computational tools to directly extract the $P_n$ functions up to $n = 8$ from multi-modal imaging data. Using simple model microstructures, we show that these novel statistical descriptors effectively "decompose" the structural features of interest into a set of "polytope basis", allowing one to easily detect any underlying symmetry or emerging features during the structural evolution. We apply these novel $P_n$ functions to quantify and model a variety of heterogeneous material systems, including particle-reinforced composites, metal-ceramic composites, concretes, porous materials; as well as the microstructural evolution in an aged lead-tin alloy. Our results indicate that the $P_n$ functions can offer a practically complete and compact set of basis for quantitative microstructure representation (QMR), for both static 3D complex microstructure and 4D microstructural evolution of a wide spectrum of heterogeneous material systems.

# 1. Introduction

Effective and accurate characterization and quantification of the complex microstructure of heterogeneous materials and microstructural evolution under external stimuli or during processing are crucial to processing optimization and advanced material design, yet are very challenging to achieve. One challenge involves the *in situ* characterization of the 3D microstructure containing features of interest on multiple length scales as well as the 4D evolution processes (3D microstructure + temporal evolution). This challenge has been partially addressed by the development and successful application of advanced non-destructive *in situ* imaging techniques, such as x-ray micro-computed tomography (µCT) [1, 2]. The second challenge involves the development of efficient mathematical frameworks and computational tools for quantitative representation, modeling and reconstruction of complex heterogeneous material microstructures and their evolution. In a recent TMS report on advanced computation and data in materials and manufacturing, quantitative microstructure representation (QMR) has been identified as a key technical knowledge gap in core knowledge areas "Coupling Simulations and Experiments" and "Digital Representation and Visualization" [3].

Recently, a variety of novel techniques have been developed to address the challenges in QMR, in particular, to devise reduced-dimension quantitative microstructure representations for heterogeneous material systems based on either complete 3D or lower dimensional imaging data set [4-24], and to devise the associated virtual material generation/reconstruction methods [25-48]. Examples of established microstructure representation schemes include random field models, statistical descriptor-based representations, and abstract image-based decompositions obtained via machine-learning, to name but a few. Among the descriptor-based representations, a recently developed framework for Hierarchical Materials Informatics, based on complete 2-point statistics and its lower dimensional projections [49-51], can directly yield accurate estimates of material properties [42, 52-59] and thus, has been incorporated into various integrated computational material design frameworks.

In our opinion, an ideal set of material representations shall acquire the following crucial properties: (1) *conciseness* to facilitate tractable sampling and inverse design within the low-dimensional representation space; (2) *expressiveness* to enable high-fidelity reconstruction of complex microstructures with heterogeneous morphologies, and accurate predictions of material properties of interest; (3) *universality* to allow efficient computation from distinct multi-modal imaging data and to enable data fusion for microstructure modeling; (4) *interpretability* to allow easy and intuitive understanding of the key morphological features in the material systems and their evolution from the representations. We note this last property "*interpretability*" is crucial for microstructural evolution representation, and that the limitation of the state-of-art frameworks, including the 2-point statistics based material informatics approach and the data-driven machine learning techniques, is mainly the lack of "interpretability". For example, the representations obtained in these frameworks (e.g., low dimensional projections of the complete 2-point statistics space or the image-based decompositions) are usually too abstract and very challenging to visualize and interpret physically. However, we emphasize that these existing representations are very effective and successful in achieving conciseness, expressiveness and universality.

Another class of widely used descriptor-based representations involves the standard n-point correlation (or probability) function $S_n$, which provides the probability of occurrence of specific n-point configurations in the material microstructure [60-63]. The complete set of $S_n$ with n = 1, 2, 3, ... ∞ provides a complete quantitative microstructure representation, and thus, determines the physical properties and performance of the material system under consideration. It has been shown that even the lower order function $S_2$ can be employed to model a wide spectrum of different heterogeneous material systems [64-71]. In practice, it is very challenging to utilize $S_n$ with n≥3, for which one needs to enumerate all distinct n-point configurations and efficiently compute and store their probability of occurrence. The resulting statistical data sets are typically much larger in size than the original imaging data for the material system. It has been shown that two-point statistics alone might not be sufficient to represent certain complex microstructures [72-75]. An alternative approach is to employ non-standard lower-order correlation functions such as the cluster functions [76, 77] or surface functions [78, 79], which encode partial higher-order n-point statistics. This method is very effective in capturing specific morphological features (e.g., clustering) [35] but cannot allow to systematically incorporate hierarchical higher order structural information for microstructure quantification.

In this paper, we develop a set of novel hierarchical statistical microstructural descriptors, called the "*n-point polytope functions*" $P_n$, which can simultaneously achieve conciseness, expressiveness, universality and interpretability, for quantitative characterization, representation and modeling of microstructural evolution during processing. These novel polytope functions successively include higher-order n-point statistics of the features of interest in the microstructure, and can be directly computed from multi-modal imaging data, including x-ray tomographic radiographs, optical/SEM/TEM micrographs, and EBSD color maps for quantification of different features of interest.

In particular, we develop highly efficient computational tools to directly extract these statistical descriptors from multi-modal imaging data, and investigate the information content of the functions for quantifying different types of structural features and their evolution. Furthermore, we apply the $P_n$ functions quantify and model a variety of heterogeneous material systems, including particle-reinforced composites, metal-ceramic composites, concretes, porous materials; as well as the microstructural evolution in an aged lead-tin alloy. Our results indicate that the $P_n$ functions can offer a practically complete and compact set of basis for quantitative microstructure representation (QMR), for both static 3D complex microstructure and 4D microstructural evolution of a wide spectrum of heterogeneous material systems.

## 2. The n-Point Polytope Functions $P_n$

### 2.1. Definition of the n-point polytope functions $P_n$

In this section, we introduce the n-point polytope functions $P_n$. Without loss of generality, we consider a heterogeneous microstructure (either in 3D or a 2D slice) in which the different structural features are segmented and grouped into different "phases". A simple example is a composite microstructure contains a "matrix phase" and a "particle phase", see Fig. 1a. Note this microstructure can be a snapshot from an evolution process. The definition of $P_n$ is then given as follows:

$P_n(r) \equiv$ *Probability that all of the n vertices of a randomly selected regular n-point polytope with edge length r fall into the phase of interest*.

Base on this definition, one can derive two sets of the $P_n$ functions. The first set involves *n-point regular polygons*, for which the vertex (edge) number *n* can take any positive integer values; and in the limit $n \to \infty$, the shape becomes a circle (in this limiting case, the quantity *r* is the radius of the circle, instead of the edge length). We note that the n-point polygon functions can be computed from both 2D slices and full 3D microstructure. The other set involves 3D polyhedra whose edges are of the same length. Only a small number of 3D polyhedra satisfy this condition, including the five *Platonic solids* (i.e., the regular polyhedra: tetrahedron, octahedron, dodecahedron, icosahedron, and cube) and the thirteen *Archimedean solids* (i.e., the semi-regular polyhedra) (see Fig. 1b) [80]. In the case of n = 2, the 2-point polytope function $P_2$ is identical to the standard 2-point correlation function $S_2$.

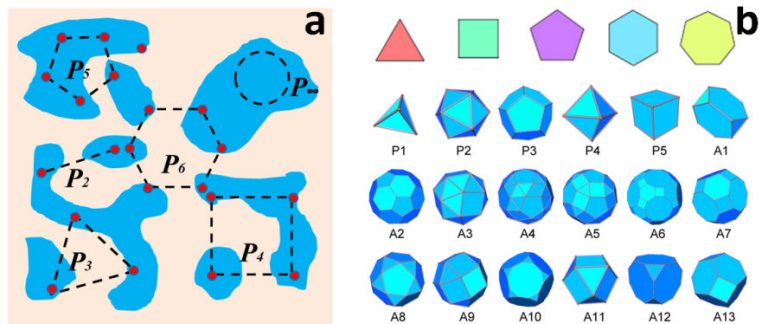

**Fig. 1:** (a) Schematic illustration of stochastic events contributing to the $P_n$ functions in the case of regular polygons. (b) Examples of regular polygons (triangle, square, pentagon, hexagon, and decagon) and polyhedra (including the Platonic solids P1-P5, and Archimedean solids A1-A13) for computing the $P_n$ functions.

Fig. 1a schematically illustrates the stochastic events that contribute to the $P_n$ functions in the case of regular polygons. For $r = 0$, the polygon reduces to a single point, and $P_n(r=0)$ gives the volume fraction $\varphi$ of the phase of interest (i.e., the probability a randomly selected point falling into the phase of interest). For finite $r$ values, $P_n(r)$ provides n-point spatial correlations in the phase (feature) of interest. For very large $r$ values (e.g., $r\to\infty$), the probabilities of finding the vertices in the phase of interest are almost independent of one another, thus, we have $P_n(r\to\infty) \approx \varphi^n$, where $\varphi$ is the volume fraction of the phase of interest in the system. These asymptotic behaviors allow us to introduce a convenient re-scaled form of the $P_n$ functions, i.e.,

$$f_n(r) = \left[P_n(r) - \varphi^n\right] / \left[\varphi - \varphi^n\right] \tag{1}$$

with $f_n(r=0) = 1$ and $f_n(r\to\infty) = 0$. Finally, we note that one can define "cross-correlation" polytope functions, by requiring a subset of the vertices falling into different phases (features) in the microstructure. In this paper, we will mainly focus on the "auto" polytope functions defined with 2D polygons, in which all of the vertices of the polygons fall into the same phase of interest.

## 2.2. Extracting $P_n$ functions from imaging data

The probability-based definition of the $P_n$ functions allow us to easily compute these function from microstructural data, including both 2D images and 3D digital representations of the material microstructure. For example, in order to compute the value of $P_n(r)$ with $r = r^*$, the following procedure is used:

**(i)** We first generate a regular n-polytope (e.g., a n-polygon) with edge length $r^*$;

**(ii)** This n-polygon is then placed in the material microstructure with randomly selected center location and random orientation, for $M$ times (see Fig. 2a);

**(iii)** Each time the polygon is placed in the system, we check whether all of its vertices fall into the phase of interest (i.e., a "success" event), and we count the total number of success event $Ms$ out of a total of $M$ trials;

**(iv)** We compute $P_n(r = r^*) = Ms/M$, which is the probability that a randomly selected n-polygon having all its vertices fall into the phase of interest.

This procedure is repeated for different $r$ values to compute the full $P_n(r)$ function.

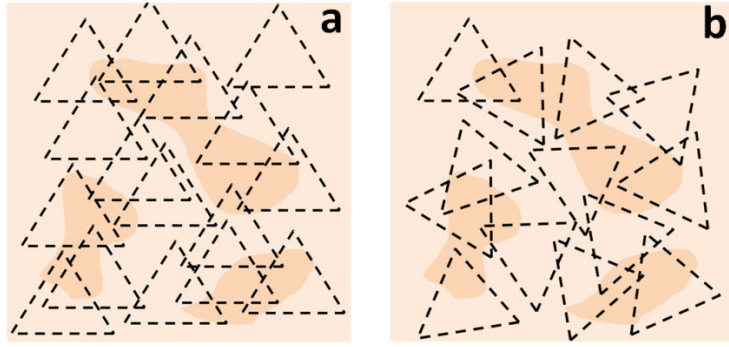

**Fig. 2:** Schematic illustration of different sampling templates for computing $P_n$ functions (in the case of n=3) from images: (a) Isotropic sampling, in which both the location and orientation of the polygon template are randomly selected. (b) Directional sampling, in which only the location of the template is randomly selected.

We note that for a digitized representation of a material microstructure (i.e., an image), a point falls into the phase of interest if it fall into a pixel (or voxel in 3D) of that phase. Therefore, the edge length of the pixel/voxel defines the smallest distance in the system and provides a natural unit for measuring the distance. In addition, besides placing the polytope template with random orientations (i.e., isotropic sampling), one can fix the orientation of the polytopes while placing them at randomly selected locations, see Fig. 2b. We refer to this later case as "directional sampling".

## 3. Information Content of $P_n$ Functions

In this section, we investigate the information content of the $P_n$ functions using simple model microstructures. As shown in Fig.3a and Fig. 4a, the model microstructures are composed of congruent equilateral triangle particles arranged on a square lattice and triangular lattice, respectively. Within each model microstructure, the orientations of all triangles are the same; and different microstructures contain triangles with different orientations.

Fig. 3b and 3c show the directional 3-point polytope functions $P_3$ for the model microstructures shown in Fig. 3a. In particular, the directional $P_3$ functions are computed by fixing the orientation of the equilateral triangular sampling templates vertically and horizontally respectively (see the insets of the figures). It can be seen that the $P_3$ functions for all microstructures start with the same value at $r = 0$, which is the volume fraction of the particles; and monotonically decrease as one moves away from $r = 0$. This behavior is well known in the case of two-point correlation function $S_2 \equiv P_3$. This is because for small $r$, the majority of contributions to the correlation functions are from the events in which the sampling templates entirely fall into a single particle. Clearly, the success of such events rapidly decreases as the size of the template increases. As we will see later, this monotonic decaying behavior is the case for all $P_n$ functions due to the same reason. We note that when the orientation of the triangular particles is consistent with the orientation of the sampling triangular template, the resulting $P_3$ function exhibits a slower decay for small $r$. This is because in such cases, the sampling templates with larger sizes (i.e., associated with larger $r$ values) can still entirely fall into a single particle.

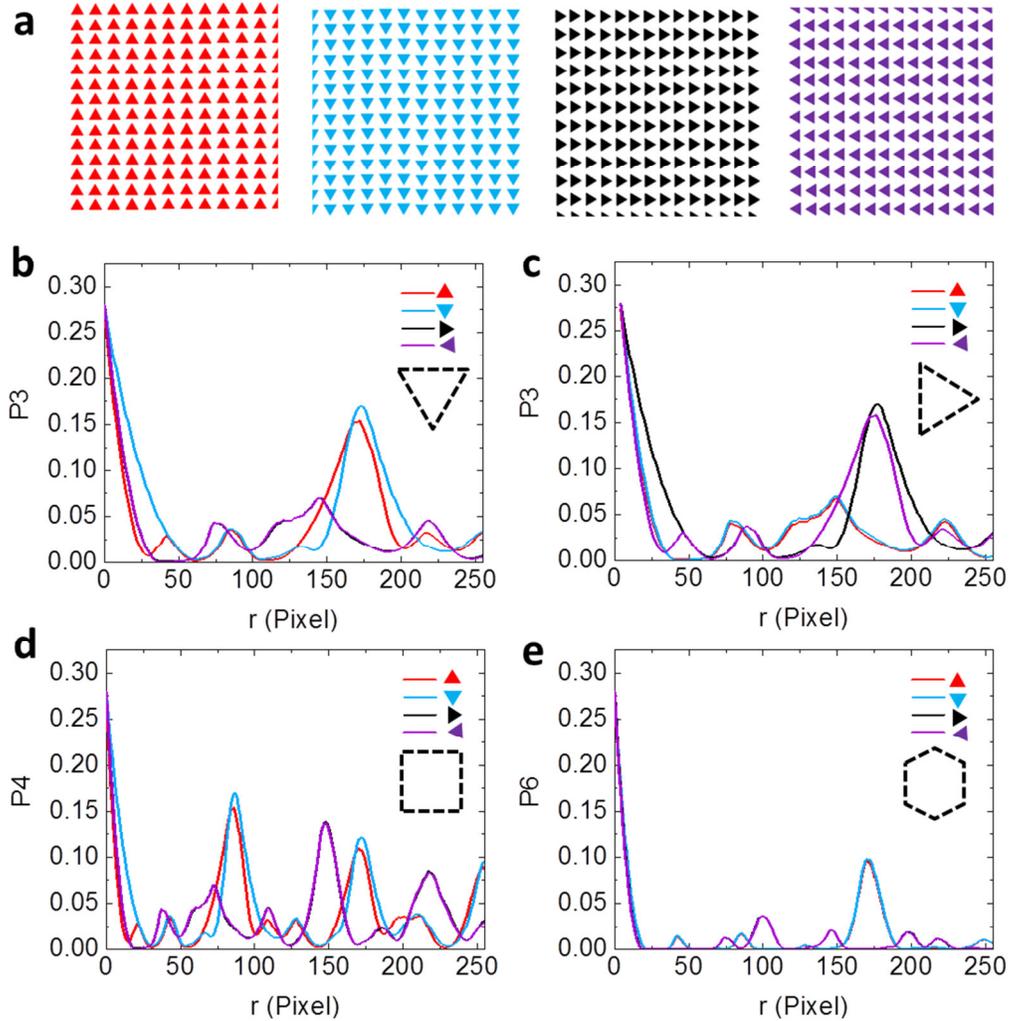

**Fig. 3**: (a) Model microstructures composed of congruent equilateral triangle particles with different orientations arranged on a square lattice. The resolution of the images is 1048 by 1048 pixels. (b) Directional 3-point polytope functions $P_3$ for the model microstructures associated with vertically oriented triangular sampling templates. (c) Directional 3-point polytope functions $P_3$ for the model microstructures associated with horizontally oriented triangular sampling templates. (d) 4-point polytope functions $P_4$ for the model microstructures. (e) 6-point polytope functions $P_6$ for the model microstructures. The unit of distance is one pixel length.

Interestingly, it can be seen that beyond the initial decay, the $P_3$ functions exhibit significant peaks at certain intermediate and event large $r$ values (e.g., $r \approx 180, 290, 340, 430$ pixels). These peaks indicate the existence strong triangular correlations (i.e., hidden triangular patterns/symmetry) on the length scales as defined by the associated $r$ values. We note it is very surprising to detect the emergence of large-scale triangular correlation (symmetry) in packing arrangements based on square lattice. This example also indicates that the novel statistical descriptors we developed are very efficient in capturing hidden order and spatial correlations in the microstructure.

Fig. 3d shows the 4-point polytope functions $P_4$ for the model microstructures. A series of peaks (i.e., oscillations) beyond the initial decay can be clearly observed. These peaks correspond to the square correlation/symmetry on different length scales in the packings, which are resulted from the square symmetry of the underlying packing lattice. The $P_4$ functions are almost identical for all four microstructures (see Fig. 3a), indicating that this function is not sensitive to the local particle shape, but only depends on the packing arrangements of the particles.

Fig. 3e shows the 6-point polytope functions $P_6$ for the model microstructures. It can be seen that the initial decay of $P_6$ is much faster than that in $P_3$ and $P_4$. This is because it is more difficult to entirely fit a hexagon in the particle compared to a triangle or a square with the same edge length. The $P_6$ functions also exhibit significant peaks on certain intermediate and large length scales (e.g., $r \approx 180, 290, 340, 430$ pixels), indicating hidden hexagonal symmetry/correlations on that length scale. These length scales are completely consistent with the length scales associated with the triangular peaks in $P_3$.

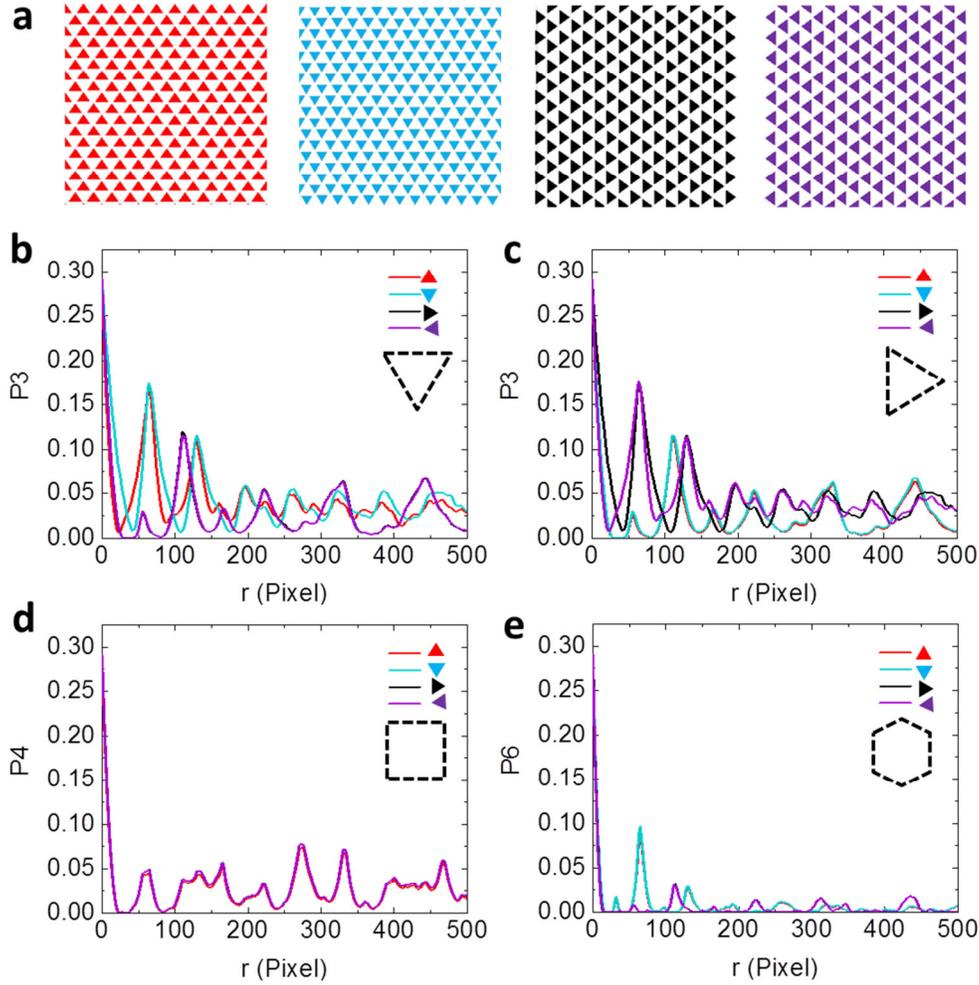

**Fig. 4**: (a) Model microstructures composed of congruent equilateral triangle particles with different orientations arranged on a triangular lattice. The resolution of the images is 1048 by 1048 pixels. (b) Directional 3-point polytope functions $P_3$ for the model microstructures associated with vertically oriented triangular sampling templates. (c) Directional 3-point polytope functions $P_3$ for the model microstructures associated with horizontally oriented triangular sampling templates. (d) 4-point polytope functions $P_4$ for the model microstructures. (e) 6-point polytope functions $P_6$ for the model microstructures. The unit of distance is one pixel length.

Fig. 4b and 4c show the directional 3-point polytope functions $P_3$ respectively associated with vertically and horizontally oriented triangular sampling templates, for the model microstructures shown in Fig. 4a. The $P_3$ functions initially decay and exhibits a series of significant peaks (oscillations) with decaying magnitude as $r$ increases. These peaks are associated with triangular correlations on different length scales, resulted from the underlying triangular packing lattice. Fig. 4d shows the 4-point polytope functions $P_4$ for the model microstructures. The hidden square correlations on large length scales (e.g., $r \approx$ 275, 340 pixels) are again picked up and manifested as the associated peaks. Fig. 4e shows the 6-point polytope functions $P_6$ for the model microstructures. Similar to the $P_3$ functions, the $P_6$ functions exhibit significant peaks and strong oscillations, indicating the hidden hexagonal correlations resulted from the triangular packing arrangements.

We can see from these examples that the $P_n$ functions can provide a systematic "decomposition" of the features of interest in the microstructure in terms of a series of polytope basis and very effective in capturing hidden symmetry/spatial orders in the system. Therefore, the set of $P_n$ functions offers a systematic way to devise more accurate microstructure representations by successively incorporating higher order morphological information.

## 4. Quantification of Complex Heterogeneous Materials and Microstructural Evolution

In this section, we employ the $P_n$ functions to quantify and model a variety of heterogeneous materials with distinct microstructures, including particle reinforced composites, bi-phase interpenetrating composites and porous materials. In addition, we also apply the $P_n$ functions to quantify and model the microstructural evolution (e.g., coarsening development) in a lead-tin alloy aged at 175°C up to 120 hours.

### 4.1. Quantification of complex microstructures

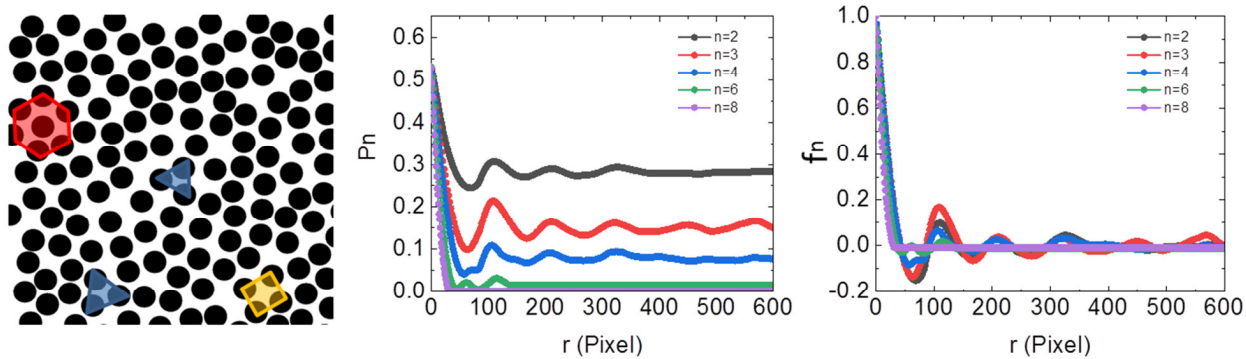

**Fig. 5:** (a) 2D microstructure composed of equal-sized hard spheres in a matrix. (b) The $P_n$ functions for the particle phase with n =2, 3, 4, 6 and 8. (c) The corresponding rescaled $f_n$ functions highlighting the spatial correlations (c.f. Eq. (1)).

Fig. 5 shows the quantification of a 2D microstructure composed of equal-sized hard spheres in a matrix [31], i.e., a packing (see Fig. 5a). The sphere packing is generated using Monte Carlo simulations [80]. The size of the image is 1024 by 1024 pixels and the diameter of the particle is $D$ = 65 pixels. Fig. 5b shows the $P_n$ functions for the particle phase with n =2, 3, 4, 6 and 8. Similar to the cases of the model microstructures discussed in Sec. 3, all $P_n$ functions initially decay from the volume fraction $\varphi$=0.545 as $r$ increases from 0. The positions of the first minimum in the $P_n$ functions for small n values roughly correspond to the linear size of the particle (~ 65 pixels). After the initial decay, all $P_n$ functions studied here except for $n = 8$ exhibit strong oscillations and the first peaks in different $P_n$ functions occur at approximately the same $r$ values. These oscillations respectively indicates strong pair, triangle, square and hexagonal correlations on different length scales in the system. Fig. 5a illustrates examples of such correlations, which are all associated with the mean nearest neighbor separate distance, i.e., the distance associated with the first peak in $P_2$. These correlations result from the tendency for the particles to self-organize on a triangular lattice at high densities. The $P_8$ is almost flat after the initial decay, indicating the system does not possess any octagonal correlations on intermediate and large length scales. Fig. 5c shows the corresponding rescaled $f_n$ functions highlighting the spatial correlations (c.f. Eq. (1)).

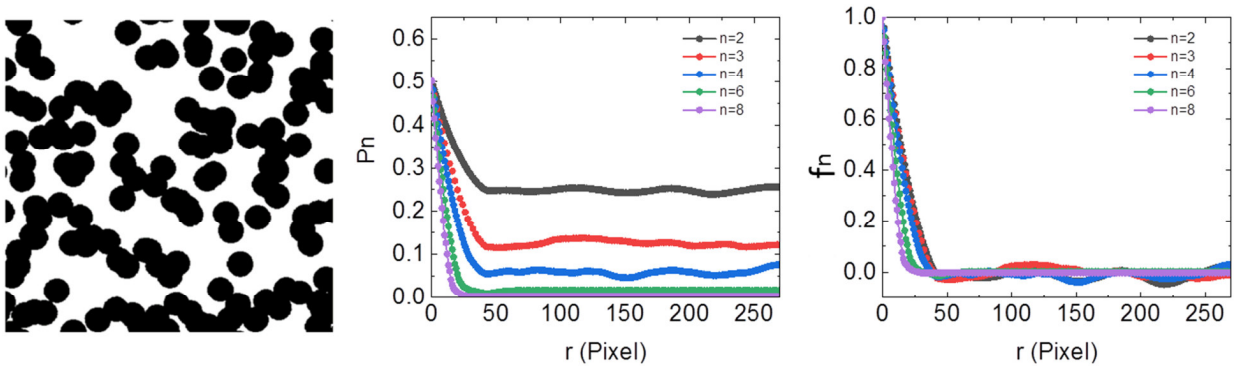

**Fig. 6:** (a) 2D microstructure composed of equal-sized overlapping spheres in a matrix. (b) The $P_n$ functions for the particle phase with n =2, 3, 4, 6 and 8. (c) The corresponding rescaled $f_n$ functions highlighting the spatial correlations (c.f. Eq. (1)).

Fig. 6 shows the quantification of a 2D microstructure composed of equal-sized overlapping spheres in a matrix [31] (see Fig. 6a). The spheres are randomly placed in the matrix without any built-in spatial correlations. The size of the image is 512 by 512 pixels and the diameter of the particle is $D = 45$ pixels. Fig. 6b shows the $P_n$ functions for the particle phase with n =2, 3, 4, 6 and 8. Similar to the previous systems, all $P_n$ functions initially decay from the volume fraction $\varphi=0.50$ as $r$ increases from 0. After the initial decay, all $P_n$ functions are virtually flat, indicating that the particles possess no spatial correlations of any symmetry on any length scales beyond the diameter of the particles. We note that for the totally random overlapping sphere system, the $P_n$ functions possess the analytical expression $P_n I = \exp[-\rho v_n(r; R)]$, where $\rho$ is the number density of the spheres in the system (i.e., number of spheres per unit volume) and $v_n(r; R)$ is the volume of the union of $n$ spheres with radius $R$ with centers placed at the vertices of a n-polytope with edge length $r$. Fig. 6c shows the corresponding rescaled $f_n$ functions highlighting the spatial correlations (c.f. Eq. (1)).

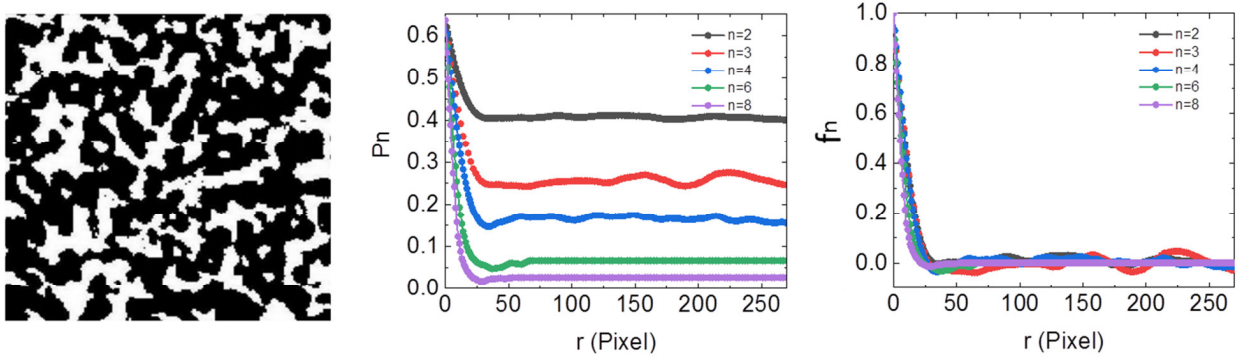

**Fig. 7:** (a) 2D image of a concrete microstructure composed of reinforcement rocks (shown in black) and the cement paste (shown in white). (b) The $P_n$ functions for the rock (black) phase with n =2, 3, 4, 6 and 8. (c) The corresponding rescaled $f_n$ functions highlighting the spatial correlations (c.f. Eq. (1)).

Fig. 7 shows the quantification of a 2D concrete microstructure composed of reinforcement rocks (shown in black) and the cement paste (shown in white) [35] (see Fig. 7a). The rock particles possess complex polygonal morphologies and a wide size distribution. The size of the image is 400 by 400 pixels. Fig. 7b shows the $P_n$ functions for the rock phase with n =2, 3, 4, 6 and 8. Similar to the previous systems, all $P_n$ functions initially decay from the volume fraction $\varphi=0.48$ as $r$ increases from 0. The $r$ value associated with the first minimum in the functions provides the average particle size in the system, i.e., ~ 40 pixels. After the initial decay, the lower-order $P_n$ functions (e.g., n≤4) exhibit weak oscillations, reflecting the spatial correlations resulted from the mutual exclusion effects of the rock particles. The correlations are much weaker compared those in hard-sphere systems, mainly due to the anisotropy and size polydispersity of the rock particles. Fig. 7c shows the corresponding rescaled $f_n$ functions highlighting the spatial correlations (c.f. Eq. (1)).

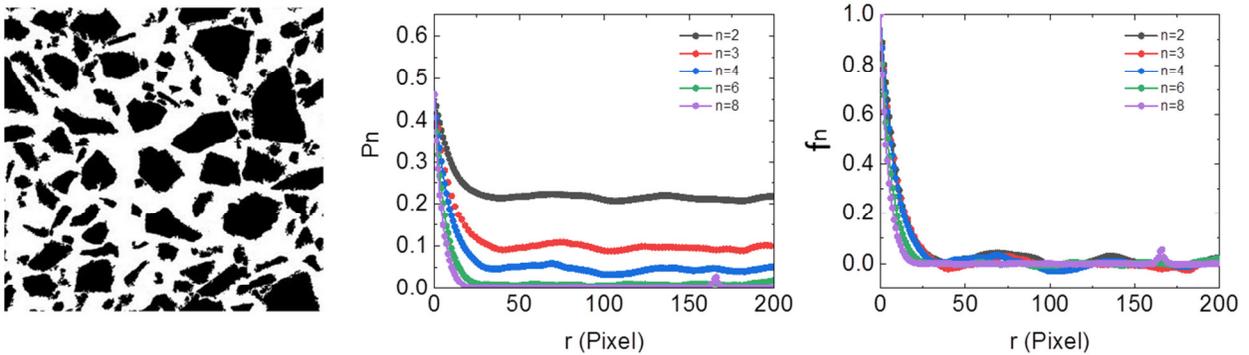

**Fig. 8:** (a) 2D image of a Fontainebleau sandstone microstructure composed of the rock phase (shown in black) and the pore phase (shown in white). (b) The $P_n$ functions for the rock (black) phase with n =2, 3, 4, 6 and 8. (c) The corresponding rescaled $f_n$ functions highlighting the spatial correlations (c.f. Eq. (1)).

Fig. 8 shows the quantification of a 2D microstructure of a Fontainebleau sandstone composed of the rock phase (shown in black) and the pore phase (shown in white) [34] (see Fig. 8a). Similar to the concrete microstructure, the rock particles possess complex shapes and a wide size distribution. In addition, the particles are densely packed and compressed such that their boundaries are fused and cannot be clearly distinguished. The size of the image is 512 by 512 pixels. Fig. 8b shows the $P_n$ functions for the rock phase with n =2, 3, 4, 6 and 8. Similar to the previous systems, all $P_n$ functions initially decay from the volume fraction $\varphi=0.82$ as $r$ increases from 0. The $r$ value associated with the first minimum in the functions provides the average particle size in the system, i.e., ~ 32 pixels. After the initial decay, the $P_n$ functions exhibit very weak oscillations for small and intermediate $r$ values. This behavior is similar to that observed for the overlapping sphere system, as the rocks are compressed and fused which can be effectively considered as "overlapped" in the boundary regions. Fig. 8c shows the corresponding rescaled $f_n$ functions highlighting the spatial correlations (c.f. Eq. (1)).

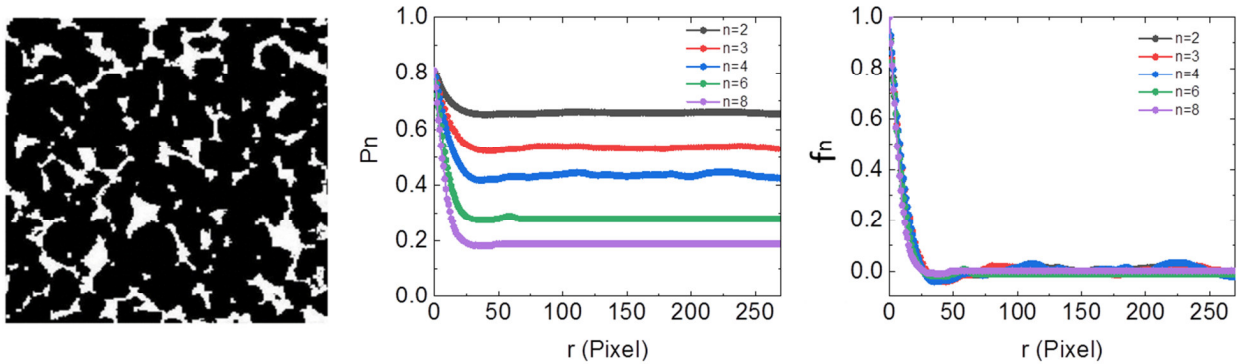

**Fig. 9:** (a) 2D image of an interpenetrating metal-ceramic composite microstructure composed of the boron-carbide phase (shown in black) and the aluminum phase (shown in white). (b) The $P_n$ functions for the boron-carbide (black) phase with n =2, 3, 4, 6 and 8. (c) The corresponding rescaled $f_n$ functions highlighting the spatial correlations (c.f. Eq. (1)).

Fig. 9 shows the quantification of a 2D an interpenetrating metal-ceramic composite composed of the boron-carbide phase (shown in black) and the aluminum phase (shown in white)[34] (see Fig. 9a). This microstructure contains "ligaments" of similar width instead of "particles". The size of the image is 512 by 512 pixels. Fig. 8b shows the $P_n$ functions for the rock phase with n =2, 3, 4, 6 and 8. Similar to the previous systems, all $P_n$ functions initially decay from the volume fraction $\varphi=0.65$ as $r$ increases from 0. The $r$ value associated with the first minimum in the functions provides the average ligament width in the system, i.e., ~ 27 pixels. After the initial decay, the $P_n$ functions exhibit weak oscillations for small and intermediate *r* values, characterizing the exclusion effects between the ligaments. Fig. 9c shows the corresponding rescaled $f_n$ functions highlighting the spatial correlations (c.f. Eq. (1)).

**4.2. Quantification of microstructure evolution**

We now employ the $P_n$ functions to quantify microstructure evolution (e.g., coarsening development) in a lead-tin alloy aged at 175°C up to 120 hours, see Fig. 10. In Ref [64], we model the evolution process using time-dependent 2-point correlation function $S_2$, which is equivalent to $P_2$. We showed that when properly scaling with the time-dependent increasing length scale (e.g., the average width of the ligaments), the $S_2$ functions correspond to different aging time all collapse approximately onto a universal curve, capturing the intrinsic density fluctuations in the system. Here, we investigate the higher order functions.

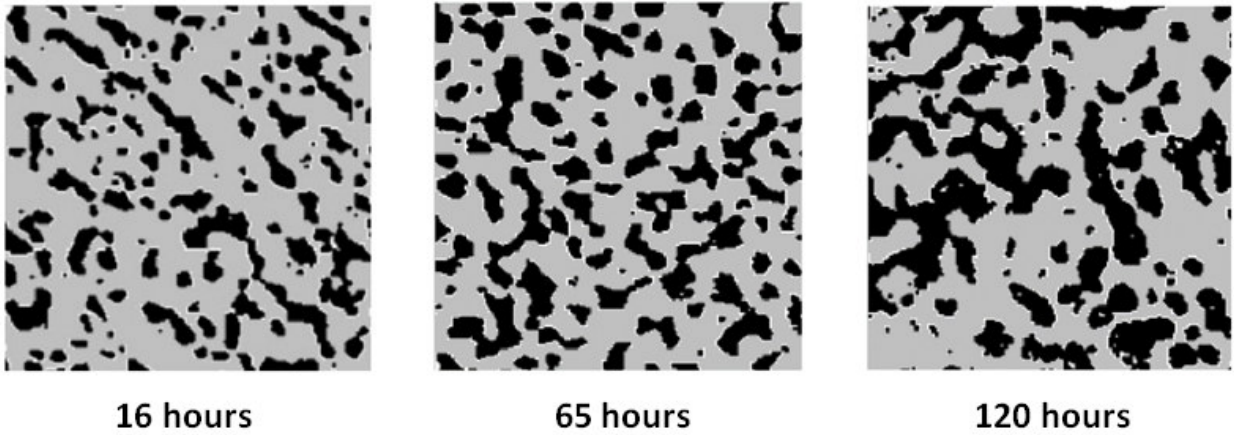

16 hours                65 hours                120 hours

**Fig. 10:** Representative microstructures of the lead-tin alloy (Pb37Sn63) at 175°C for different times. The dark region represents lead-rich phase and white region represents tin-rich phase. The linear size of the microstructure is ~ 100 μm. The resolution of the image is 600 by 600 pixels.

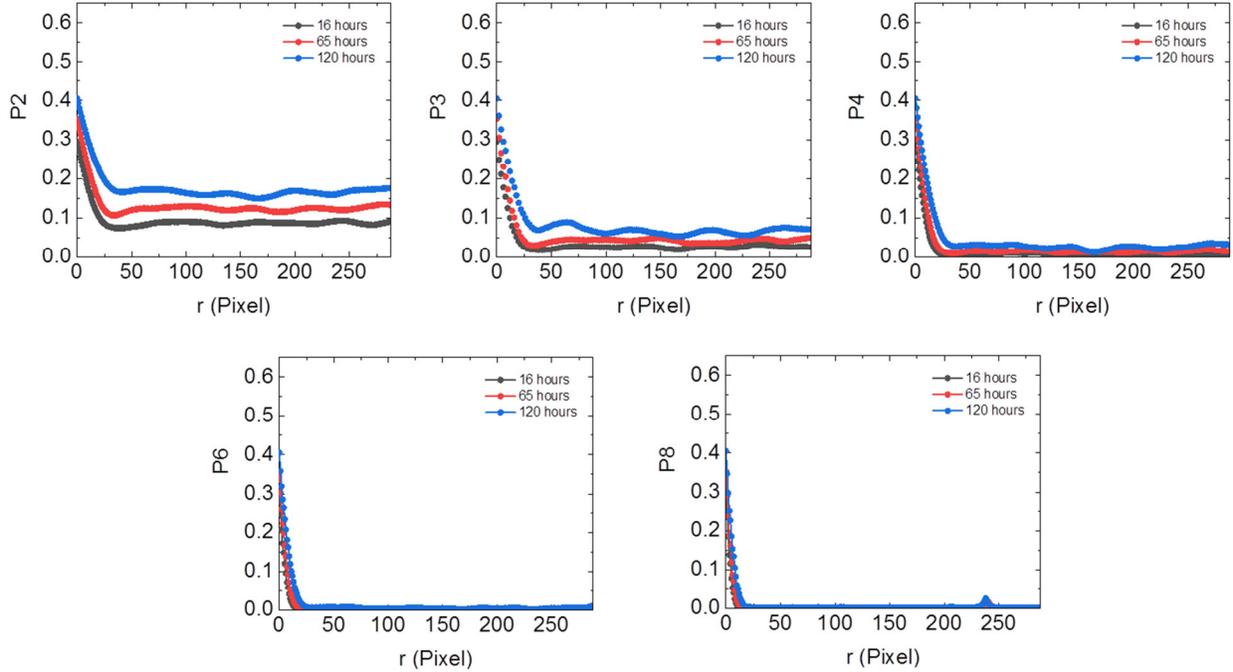

**Fig. 11:** The $P_n$ functions associated with the Pb-Sn alloy microstructures at different aging times.

Fig. 11 shows the $P_n$ functions (with $n$ = 2, 3, 4, 6 and 8) associated with the Pb-Sn alloy microstructures at different aging times as shown in Fig. 10. It can be seen that for $P_n$ functions, the initial decay as $r$ increases from 0 becomes slower as aging time increases. In other words, the $r$ value associated with the first minimum in the functions increases as aging time increases. As discussed above, this $r$ value characterizes the average width of the "ligaments" in the microstructure, which coarsens as aging time increases.

Moreover, at large aging time (e.g., t = 120 hours), we can see that significant triangle and square correlations emerge in the microstructure, which are manifested as the relatively strong oscillations in the corresponding $P_3$ and $P_4$ functions. Such higher order correlations are not observed for small aging time and might be resulted from the additional correlation induced due to coarsening development. We note that the $P_2$ functions also exhibit stronger oscillations as coarsening proceeds. It is now clear that the observed pair correlations are associated with the emergence of higher order correlations due to coarsening. This example illustrates the importance of incorporating higher-order statistics for accurate quantification and representation of complex microstructural evolution.

## 5. Conclusions and Discussion

In this paper, we have introduced a novel set of hierarchical statistical descriptors, i.e., the n-point polytope functions $P_n$ for quantitative representation of complex heterogeneous materials and microstructure evolution. The $P_n$ functions provide the probability of finding regular n-polytopes with vary sizes in the material phase of interest, which are a subset of the full n-point correlation function $S_n$. We have shown that the $P_n$ functions allow one to systematically incorporate higher order spatial correlations for microstructure quantification in a concise, expressive, interpretable and universal fashion, and provide an effective decomposition of the features of interest into polytope correlations. The utility of the $P_n$ functions has been demonstrated by applying them to quantify and model a variety of complex heterogeneous material systems as well as microstructural evolution. Our results suggest that the n-point polytope functions offer a complete and compact set of basis for quantitative microstructure representation (QMR), for both static 3D complex microstructure and 4D microstructural evolution of a wide spectrum of heterogeneous material systems.

In addition, we note that the polytope functions can be readily incorporated into material reconstruction algorithms for fast virtual 3D microstructure generation and also allow fast material property prediction via analytical structure-property mappings. The reconstruction can be either formulated as a stochastic optimization [31, 34, 35] or constrained neural network models [81, 82]. This would potentially enable in-processing material property monitoring and real-time processing optimization.

**Acknowledgement**

This work is supported by ACS Petroleum Research Fund under Grant No. 56474-DNI10 (Program manager: Dr. Burtrand Lee). Y.J. and N.C. acknowledge the Office of Knowledge Enterprise Development (OKED) and the Fulton Schools of Engineering for a seed grant.